\documentclass[11pt]{article}
\usepackage{blois,epsfig}
\usepackage{multirow}

\def\Journal#1#2#3#4{{#1} {\bf #2}, #3 (#4)}

\def\aap{A\&A}%
\def\apj{ApJ}%

\def\be{\begin{equation}}
\def\ee{\end{equation}}
\def\bea{\begin{eqnarray}}
\def\eea{\end{eqnarray}}

\def\degr{^{\circ}}

\begin{document}
\vspace*{4cm}
\title{Search for Galactic Cosmic Ray Sources with H.E.S.S.}

\author{Nukri Komin on behalf of the H.E.S.S. collaboration }

\address{LAPP, CNRS/IN2P3, 9 Chemin de Bellevue,\\
74941 Annecy-le-Vieux, France }

\maketitle\abstracts{
Supernova remnants (SNRs) are the prime candidates for the acceleration of the Galactic Cosmic Rays. Tracers for interactions of Cosmic Rays with ambient material are gamma rays at TeV energies, which can be observed with ground based Cherenkov telescopes like H.E.S.S. In the recent years H.E.S.S. has detected several SNRs and interactions of SNRs with molecular clouds. Here the current results of these observations are presented and possible leptonic and hadronic scenarios are discussed. It is shown that it is likely that SNRs are the sources of Galactic Cosmic Rays.
}

\section{Introduction}

Since the discovery of the Cosmic Rays in 1912 their origin remains a mystery. Very promising candidates for acceleration of at least the Galactic Cosmic Rays (up to $\approx 10^{15}\,\mathrm{eV}$) are SNRs; it has been shown \cite{Ginzburg} that several percent of the energy of every supernova explosion (of the order of $10^{51}\,\mathrm{erg}$) can easily supply the entire power found in the Cosmic Rays. The theory of Diffusive Shock Acceleration (DSA) \cite{DSA} shows that particles can be accelerated up to relativistic energies in shock fronts. Among others, these shock fronts can be found in the shells of SNRs.

As Cosmic Rays as charged particles are deflected in magnetic fields their direct observation does not carry any information on their origin. Observations of photons produced by relativistic particles are much more promising messengers. The discovery of synchrotron emission from SNRs shows that at least electrons and positrons are accelerated in the shells of SNRs. A tracer for acceleration of protons are gamma rays from hadronic interactions of relativistic protons with ambient material. The resulting photon spectrum is relatively flat and ranges from MeV to TeV energies. These gamma-rays can be observed by satellite experiments in the MeV and GeV energy range (e.g. by the \textit{Fermi}/LAT) or with ground based telescopes above $\approx 100\,\mathrm{GeV}$.

The High Energy Stereoscopic System (H.E.S.S.) is a Cherenkov telescope array located in Namibia for the observation of very heigh energy (VHE) gamma rays ($\approx 100,\mathrm{GeV}$ up to several tens of TeV) \cite{Crab}. Due to its location in the southern hemisphere it can observe a large part of the Galactic plane. Stereoscopic observations allow a reconstruction of the photon direction with an angular resolution of better than $0.1\degr$. The photon energies can be reconstructed with an accuracy of 25\%.

\section{Observations of Supernova Remnants}

Since the start of observations with H.E.S.S. 5 shell-type SNRs have been detected in VHE gamma rays. Four SNRs were known emitters of non-thermal X-rays (RX\,J1713.7$-$3946, \\ RX\,J0852$-$4622, RCW\,86, SN\,1006; one SNR (HESS J1731$-$347) was discovered in the survey of the Galactic plane conducted with HESS \cite{UnId} and was subsequently discovered in radio and X-rays.


For the SNRs RX\,J1713.7$-$3946, RX\,J0852$-$4622 and HESS J1731$-$347 the TeV morphology is clearly a shell-like structure, which is statistically preferred over a centre-filled morphology.
The morphology of SN\,1006 are of two spots which are consistent with bright spots of non-thermal X-ray emission on the SNRs shell. For RCW\,86 a shell-like morphology could not yet been confirmed being statistically significant, further observations may clarify the morphology. The first part of Table~\ref{tab:SNRspec} lists the SNRs detected with H.E.S.S. with their references and morphological and spectral parameters. The diameters range from $0.5\degr$ up to $2\degr$. The energy spectra follow straight power laws, only for RX\,J1713.7$-$3946 an exponential cut-off around 18\,TeV has been detected. The spectral indices are between 2.0 and 2.5, which is consistent with predictions from DSA. The detection of gamma-rays from SNRs is evidence for particle acceleration in the SNR shells.

\begin{table}[t]
\caption{Properties of the TeV emission of the shell-type supernova remnants (first part of the table) and SNR/Molecular Cloud interactions (second part of the table) detected with H.E.S.S.\label{tab:SNRspec}}
\vspace{0.4cm}
\begin{center}
\setlength{\tabcolsep}{4pt}
\begin{tabular}{|l|rrrrr|}
\hline
name 	& diam.	
		& spectral index & diff. flux 			 
		& integral flux  
		& energy flux 	 \\

		&			& 				 
		& $\Phi(1\,\mathrm{TeV})$ [$10^{-8}$ 
		& $I_\gamma(1-10\mathrm{TeV})$	 
		& $F_\gamma(1-10\mathrm{TeV})$			\\
		&			& 				 
		& $\mathrm{TeV}^{-1}\mathrm{m}^{-2}\mathrm{s}^{-1}$]	
		& [$\mathrm{m}^{-2}\mathrm{s}^{-1}$] 
		& [$\mathrm{erg\,cm}^{-2}\mathrm{s}^{-1}$] \\
\hline

RX\,J1713.7$-$3946 \cite{J1713} 	& $1.2\degr$ 
					& $2.04 \pm 0.04$ \footnotemark[1]   
					& $21.3\pm0.5$ 
					& $2.63 \,\, 10^{-7}$ 
					& $7.5\,\,10^{-11}$	 
					\\

RX\,J0852$-$4622 \cite{Junior}	& $2\degr$ 
					& $2.24 \pm 0.04$
					& $18.8 \pm 0.8$ 
					& $1.4\,\, 10^{-7}$ 
					& $5.3\,\, 10^{-11}$
					\\

RCW\,86 \cite{RCW86}				& $0.9\degr$
					& $2.54 \pm 0.12$
					& $3.6 \pm 0.5$ 
					& $2.35 \,\, 10^{-8}$ 
					& $8.5 \,\, 10^{-12}$
					\\
					
SN\,1006 \cite{SN1006}			& $0.48\degr$
					& $\left\{ \begin{array}{c} 2.35 \pm 0.14 \\ 2.29 \pm 0.18 \end{array} \right. $
					& $\left. \begin{array}{c} 0.23 \pm 0.04 \\ 0.16 \pm 0.04 \end{array} \right\}$ 
					& $3.8 \,\, 10^{-9}$
					& $8.3 \,\, 10^{-13}$
					\\

HESS\,J1731$-$347 \cite{J1731}	& $0.54\degr$
					& $2.32 \pm 0.06$
					& $4.67 \pm 0.19$ \footnotemark[2] 
					& $1.9 \,\, 10^{-8}$
					& $6.9 \,\, 10^{-12}$
					\\
\hline
HESS\,J1745$-$303 \cite{J1745}	& $\approx 0.4\degr$
					& $2.71 \pm 0.11$
					& $2.84 \pm 0.23$
					& $1.6\,\, 10^{-8}$
					& $5.1 \,\, 10^{-12}$
					\\
					
W28 \cite{W28}		& mult.
					& $\left\{ \begin{array}{c} 2.49 \pm 0.14 \\ 2.66 \pm 0.27 \end{array} \right. $
					& $\left. \begin{array}{c} 1.86 \pm 0.19 \\ 0.75 \pm 0.11 \end{array} \right\}$
					& $1.64 \,\, 10^{-8}$
					& $3.5 \,\, 10^{-12}$
					\\
					
CTB\,37A \cite{CTB37A}	& ?
					& $2.30 \pm 0.13$
					& $0.87 \pm 0.10$
					& $6.4\,\, 10^{-9}$
					& $2.3 \,\, 10^{-12}$
					\\
					
W\,51C \cite{W51C}	& $\approx 0.2\degr$
					& -
					& -
					& $0.62\,\, 10^{-8}$ \footnotemark[3]
					& -
					\\

\hline
\end{tabular}
\end{center}
\end{table}
\footnotetext[1]{exponential cut-off at $17.9 \pm 3.3\,\mathrm{TeV}$}
\footnotetext[2]{at $0.783\,\mathrm{TeV}$}
\footnotetext[3]{currently only the integral flux is published}

Another proof of particle acceleration in SNRs are the detection of gamma rays from the interactions of SNRs with molecular clouds (MCs). Relativistic protons escape the SNR and interact with the material in a nearby molecular cloud. The high density in the cloud enhances the gamma-ray emission, and the gamma-ray morphology is in general correlated with the morphology of the MCs as traced by radio observations of CO emission lines. The second part of Table~\ref{tab:SNRspec} summarises the objects detected with H.E.S.S. The morphologies are rather irregular: the extension of CTB\,37A is not statistically significant, several clouds are seen around W\,28, and the diameters of the other sources are several tenth of a degree. The spectra follow straight power laws but are in general softer than those seen from shell-type SNRs. The spectral indices can go up to 2.7, similar to the index of the Cosmic Rays measured directly.

\section{Interpretation}

The fact that the gamma-ray emission from MCs is correlated with the distribution of the target material for hadronic interaction and the absence of non-thermal X-ray emission from these sources is a good indication that the emission is indeed from relativistic protons. It can be shown that the Cosmic Ray density is enhanced with respect to the density in the solar neighbourhood. And the nearby SNR which is in interaction with the MC is a very good candidate for being the sources of these Cosmic Rays.

\begin{table}[t]
\caption{Leptonic and hadronic interpretation of the emission of shell-type supernova remnants.\label{tab:SNRint}}
\vspace{0.4cm}
\begin{center}
\setlength{\tabcolsep}{5pt}
\begin{tabular}{|l|rrrr|}
\hline
name 	& X-ray flux $F_X $	
		& \multicolumn{2}{c}{magnetic field} 		
		& total energy in protons 
		\\

		& $(0.5	-10\,\mathrm{keV})$	
		& leptonic	
		& shock front 
		& hadronic scenario \\

		& [$\mathrm{erg\,cm}^{-2}\mathrm{s}^{-1}$] 
		& $B_\mathrm{lep}$
		&
		& in fractions of $E_\mathrm{SNR}\approx 10^{51}\mathrm{erg}$
		\\
		
\hline
RX\,J1713.7$-$3946 	& $5.5\,\,10^{-10}$ \footnotemark[4] \cite{Acero2009} 
				 	& $\approx 10 \mu\mathrm{G}$ 
				 	& $\geq65\mu\mathrm{G}$ \cite{BV2006} 
				 	& $0.1...0.3 \left( \frac{d}{\mathrm{1\,kpc}} \right)^2 \left(\frac{n}{\mathrm{1\,cm^{-3}}} \right)^{-1}$ 
				 	\\
				 	
RX\,J0852$-$4622	& $9.9\,\,10^{-11}$ \cite{Slane2001}
					& $\approx 6 \mu\mathrm{G}$ 
					& $270\mu\mathrm{G}$  \cite{Katsuda2008}
					& $0.08 \left( \frac{d}{\mathrm{750\,pc}} \right)^2 \left(\frac{n}{\mathrm{1\,cm^{-3}}} \right)^{-1}$ 
					\\

\multirow{2}{*}{RCW\,86}				
		& \multirow{2}{*}{$2.1\,\,10^{-10}$ \footnotemark[5] \cite{RCW86} }
					& \multirow{2}{*}{$15...30\mu\mathrm{G}$ }
					& $24 \mu\mathrm{G}$ \cite{Vink2006} 
						 
					& \multirow{2}{*}{$0.2...0.4 \left( \frac{d}{\mathrm{2.5\,kpc}} \right)^2 \left(\frac{n}{\mathrm{0.7\,cm^{-3}}} \right)^{-1}$ } \\
& & & $\approx 100\mu\mathrm{G}$ \cite{Voelk2005} & \\

SN\,1006			& $1.1\,\,10^{-10}$ \footnotemark[6] \cite{SN1006}
					& $\approx 30\mu\mathrm{G}$ 
					& $\geq120\mu\mathrm{G}$ \cite{Voelk2005}
					& $0.2\left( \frac{d}{\mathrm{2.2\,kpc}} \right)^2 \left(\frac{n}{\mathrm{0.085\,cm^{-3}}} \right)^{-1}$ 
					\\

HESS\,J1731$-$347	& $3.7\,\,10^{-11}$ \footnotemark[7] \cite{J1731}
					& $25\mu\mathrm{G}$
					& -
					& $0.2\left( \frac{d}{\mathrm{3.2\,kpc}} \right)^2 \left(\frac{n}{\mathrm{1\,cm^{-3}}} \right)^{-1}$ 
					\\
\hline
\end{tabular}
\end{center}
\end{table}
\footnotetext[4]{$1-10\,\mathrm{keV}$} 
\footnotetext[5]{$0.7-10\,\mathrm{keV}$} 
\footnotetext[6]{$0.5-10\,\mathrm{keV}$} 
\footnotetext[7]{for roughly $1/3$ of the shell} 

For the shell-type SNRs the situation is less clear as the synchrotron emission has to be taken into account. Synchrotron emission from the shells is evidence for a population of relativistic electrons. These electrons produce gamma-ray emission in inverse Compton (IC) scattering off the Cosmic Microwave Background (CMB) and the observed gamma-ray emission could be entirely due to IC emission. Assuming that synchrotron and IC emission are produced by the same electron population and that the gamma-ray emission is entirely due to IC emission off the CMB the magnetic field in the SNR shell can be estimated from the gamma-ray energy flux $F_\gamma$ and X-ray energy flux $F_X$ as
$
B_\mathrm{lep} = \sqrt{ 10 \frac{F_X}{F_\gamma} } \,\,\mu\mathrm{G}.
$
As shown in Table~\ref{tab:SNRint}, the magnetic fields for the shell-type SNRs estimated solely from the X-ray and gamma-ray fluxes are between 6 and 30 $\mu$G. 
A method to measure the magnetic field in a shock front is from the width of very thin filaments observed in X-rays\cite{Voelk2005}. The magnetic fields estimated from the shock front filaments are found to be of the order of several 100 $\mu$G, significantly larger than what was estimated for a purely leptonic scenario and consistent with expectations for magnetic field amplification as a result of very efficient acceleration of nuclear Cosmic Rays\cite{Voelk2005}. A higher magnetic field renders the synchrotron emission more efficient and the observed X-ray emission can be produced by a smaller electron population. The IC emission is therefore fainter and the gamma-ray radiation can be attributed to a possible hadronic scenario. 
Depending on the distance to the SNR and the density of the ambient material and assuming a certain spectral shape of the underlying proton population (typically a straight power law with index 2 between 1 GeV and 100 TeV) the total energy in protons can be estimated. This estimation for the SNRs detected by H.E.S.S. is summarised in Table~\ref{tab:SNRint} for typical values of the distance $d$ and the density $n$ of the ambient material. It can be seen that the total energy in protons is of the order of several tens of percent of an assumed supernova explosion energy of $10^{51}\,\mathrm{erg}$.

\section{Conclusion}

Observations with H.E.S.S. have revealed VHE gamma-ray emission from 5 shell-type SNRs and 4 interactions of SNRs with molecular clouds. In the case of the SNR/MC interactions the good correlation with the distribution of ambient material and the absence of synchrotron emission id a very good indication for hadronic interactions. The observed magnetic field amplification in the shells of supernova remnants seem to contradict a purely leptonic scenario and favour a hadronic scenario. SNRs are therefore very good candidates for acceleration of the Galactic Cosmic Rays. The extension of H.E.S.S. with a fifth telescope currently under construction will lower the energy threshold, allowing better studies of the energy spectra of the SNRs. Future instruments like the Cherenkov Telescope Array (CTA) will have a better angular resolution and a ten times better sensitivity, enabling detailed morphological studies and surveys for SNRs throughout the Galaxy.


\section*{Acknowledgments}
The support of the Namibian authorities and of the University of Namibia
in facilitating the construction and operation of H.E.S.S. is gratefully
acknowledged, as is the support by the German Ministry for Education and
Research (BMBF), the Max Planck Society, the French Ministry for Research,
the CNRS-IN2P3 and the Astroparticle Interdisciplinary Programme of the
CNRS, the U.K. Science and Technology Facilities Council (STFC),
the IPNP of the Charles University, the Polish Ministry of Science and 
Higher Education, the South African Department of
Science and Technology and National Research Foundation, and by the
University of Namibia. We appreciate the excellent work of the technical
support staff in Berlin, Durham, Hamburg, Heidelberg, Palaiseau, Paris,
Saclay, and in Namibia in the construction and operation of the
equipment.

\section*{References}




\begin{thebibliography}{99}
\bibitem{Ginzburg} Ginzburg \& Syrovatskii, \Journal{The Origin of Cosmic Rays}{New York}{Macmillan}{1964}.
\bibitem{DSA} Malkov \& Drury,  \Journal{Rep. Prog. Phys.}{64}{429}{2001}.
\bibitem{Crab} HESS collaboration, \Journal{\aap}{457}{899}{2006}.
\bibitem{UnId} HESS collaboration, \Journal{\aap}{477}{353}{2008}.
\bibitem{J1713} HESS collaboration, \Journal{\aap}{464}{235}{2007}.
\bibitem{Junior} HESS collaboration, \Journal{\apj}{661}{236}{2007}.
\bibitem{RCW86} HESS collaboration, \Journal{\apj}{692}{1500}{2009}.
\bibitem{SN1006} HESS collaboration, \Journal{\aap}{516}{A62}{2010}.
\bibitem{J1731} HESS collaboration, \Journal{\aap}{531}{A81}{2011}.
\bibitem{J1745} HESS collaboration, \Journal{\aap}{483}{509}{2008}.
\bibitem{W28} HESS collaboration, \Journal{\aap}{481}{401}{2008}.
\bibitem{CTB37A} HESS collaboration, \Journal{\aap}{490}{685}{2008}.
\bibitem{W51C} Fiasson for the HESS collaboration, \Journal{ICRC}{}{}{2009}.
\bibitem{Acero2009} Acero {\it et al}, \Journal{\aap}{505}{157}{2009}.
\bibitem{BV2006} Berezhko\&V\"olk,  \Journal{\aap}{451}{981}{2006}.
\bibitem{Slane2001} Slane {\it et al},  \Journal{\apj}{548}{814}{2001}.
\bibitem{Katsuda2008} Katsuda {\it et al},  \Journal{\apj}{678}{L35}{2008}.
\bibitem{Vink2006} Vink {\it et al},  \Journal{\apj}{648}{L33}{2006}.
\bibitem{Voelk2005} V\"olk {\it et al},  \Journal{\aap}{433}{229}{2005}.

\end{thebibliography}
\end{document}